\font\bbf=cmbx12

\centerline{\bbf Advances in the Proposed Electromagnetic Zero-Point Field Theory of
Inertia}
\bigskip
\centerline{Bernhard Haisch}
\centerline{Solar \& Astrophysics Laboratory, Lockheed Martin}
\centerline{3251 Hanover St., Palo Alto, CA 94304}
\centerline{E-mail: haisch@starspot.com}
\bigskip
\centerline{Alfonso Rueda}
\centerline{Dept. of Electrical Engineering and Dept. of Physics \& Astronomy}
\centerline{California State Univ., Long Beach, CA 90840}
\centerline{E-mail: arueda@csulb.edu}
\bigskip
\centerline{H. E. Puthoff}
\centerline{Institute for Advanced Studies at Austin}
\centerline{4030 Braker Lane, Suite 300, Austin, TX 78759}
\centerline{E-mail: puthoff@aol.com}
\bigskip
\centerline{\it Revised version of invited presentation at}
\centerline{34th AIAA/ASME/SAE/ASEE Joint Propulsion Conference and Exhibit}
\centerline{July 13--15, 1998, Cleveland, Ohio}
\centerline{AIAA paper 98-3143}

\bigskip
\centerline{\bf{ABSTRACT}}
\footnote{\phantom{1}}{Copyright $\copyright$ 1998 by the American Institute of Aeronautics
and Astronautics, Inc. All rights reserved.}

A NASA-funded research effort has
been underway at the Lockheed Martin Advanced Technology Center in Palo Alto
and at California State University in Long Beach to develop and test a
recently published
theory that Newton's equation of motion can be derived from Maxwell's
equations of
electrodynamics as applied to the zero-point field (ZPF) of the quantum vacuum.
In this ZPF-inertia theory,
mass is postulated to be not an intrinsic property of matter but rather a
kind of
electromagnetic drag force (which temporarily is a place holder for a more
general vacuum quantum fields reaction effect) that proves to be acceleration dependent by
virtue of the spectral
characteristics of the ZPF. The theory proposes that interactions between
the ZPF and
matter take place at the level of quarks and electrons, hence would account
for the mass of
a composite neutral particle such as the neutron. An effort to generalize
the exploratory
study of Haisch, Rueda and Puthoff (1994) into a proper relativistic
formulation has been
successful. Moreover the principle of equivalence implies that in this view
gravitation
would also be an effect originated in the quantum vacuum along the lines proposed by
Sakharov (1968). With
regard to exotic propulsion we can definitively rule out one speculatively
hypothesized
mechanism: matter possessing negative inertial mass, a concept originated
by Bondi (1957)
is shown to be logically impossible. On the other hand, the linked
ZPF-inertia and ZPF-gravity concepts open the conceptual possibility of
manipulation of
inertia and gravitation, since both are postulated to be vacuum
phenomena.
It is hoped that this will someday translate into actual technological
potential,
especially with respect to spacecraft propulsion and future interstellar travel
capability. A key question is whether the proposed ZPF-matter interactions
generating the
phenomenon of mass might involve one or more resonances. This is presently
under
investigation.

\bigskip
\centerline{\bf INTRODUCTION}
\bigskip\noindent
In an article in {\it New Scientist} science writer Robert Matthews 
(1995)
summarizes the
predictions of various scientists: ``Many researchers see the vacuum as a
central
ingredient of 21st century physics.'' The reason for this is that, despite
its name, the
vacuum is in fact far from empty. Create a perfect vacuum, devoid of all
matter and
containing not a single (stable) particle, and that region of seemingly
empty space will
actually be a seething quantum sea of activity. Heisenberg's uncertainty
relations allow
subatomic particles to flicker in and out of existence. Similar quantum
processes apply to
electromagnetic fields, and that is the origin of the electromagnetic
zero-point field
(ZPF). The entire Universe is filled with a quantum sea of electromagnetic
zero-point
energy whose properties are the basis of Matthew's predictive statement.

\bigskip\noindent
In 1994 we published an analysis which proposed that the most
fundamental property of matter --- inertia --- could be explained as an
electromagnetic
force traceable to the ZPF (Haisch, Rueda and Puthoff 1994; HRP). The
exploratory approach
we used had two weaknesses: (1) the mathematical development was quite
complex, and (2) the
calculations were dependent upon a simplified model to represent the
interactions between
material objects and the ZPF. But in spite of these two limitations, our
analysis yielded a
remarkable and unexpected result: that Newton's equation of motion, ${\bf
f}=m{\bf a}$,
regarded since 1687 as a postulate of physics, could be derived from
Maxwell's laws
of electrodynamics as applied to the ZPF. The implication is that inertia
is not an
innate property of matter, rather it is an electromagnetically-derived
force (or quantum vacuum derived force in a future more general derivation). If this
proves to be true, the potential exists for revolutionary technologies
since the
manipulation of electromagnetic phenomena is the basis of most modern
technology.
In particular, the manipulation of the vacuum electromagnetic fields is today the subject of
(vacuum) cavity quantum electrodynamics.

\bigskip\noindent
Thanks in part to a NASA research grant, we have made progress in
strengthening the basis of the ZPF-inertia hypothesis. We have been able to
rederive the
ZPF-inertia connection in a way that is mathematically much more
straightforward, that is
not dependent upon the original simplified matter-ZPF interaction model,
and that ---
importantly --- proves to be relativistic (Rueda \& Haisch 1998a, 
1998b).
This increases
our confidence considerably in the validity of the ZPF-inertia hypothesis.

\bigskip\noindent
We suggest that a change in paradigm regarding our conception of matter is
not far off. If inertia proves to be at least in part an electromagnetic force arising from
interactions
between quarks and electrons and the ZPF, this will do away with the
concept of inertial
mass as a fundamental property of matter.
\footnote{$^a$}{Vigier (1995), a former collaborator of Bohm
and de Broglie, recently proposed that the Dirac vacuum (that vast sea of virtual
electrons and positrons in the vacuum strongly coupled to the ZPF) also contributes to
inertia. We have plans to jointly explore this idea in an extension of our original
approach.} The principle of equivalence then implies that gravitational mass will need to
undergo an analogous reinterpretation. A foundation for this was laid already 30 years ago
by Sakharov (1968).
\bigskip\noindent
Lastly, the Einstein $E=mc^2$
relationship between mass and energy will also be cast in a different
light. As it now
stands this formula seems to state that one kind of ``thing,'' namely
energy, can
mysteriously be transformed into a totally different kind of ``thing,''
namely mass\dots
and vice versa. It is proposed instead that the $E=mc^2$ relationship is a
statement about
the kinetic energy that the ZPF fluctuations induce on the quarks and electrons
constituting matter (Puthoff 1989a). We are used to interpreting this
concentration of
energy associated with material objects as mass, but in fact this is more a
matter of
bookkeeping than physics. Indeed the concept of mass itself in all its
guises (inertial,
gravitational and as relativistic rest mass) appears to be a bookkeeping
convenience.
All we ever experience is the presence of a certain amount of energy or the
presence of
certain forces. We traditionally account for these energies and forces in
terms of mass,
but that appears now to be unnecessary. Interactions of the ZPF with quarks
and electrons
are what physically underlie all these apparent manifestations of mass.
This opens new
possibilities.

\bigskip\noindent
Only fifty years ago the concept of space travel was regarded by most,
including scientists
(who should have known better),  as science fiction: this in spite of the
fact that the
basic knowledge was already in place. Details and technicalities, of
course, were lacking,
but the chief handicap was --- more than anything --- a mindset that such
things simply had
to be impossible. Similar prejudices had been at work fifty years prior to
that regarding
flight. We have come to a new millenium and the first glimmerings of how to
go about
finding a way to achieve interstellar travel have started to appear on the
horizon. A very
modest --- in terms of cost --- but intellectually ambitious program has
been established
by NASA: The {\it Breakthrough Propulsion Physics Program} (BPP).
The rationale is stated as follows:
\footnote{$^b$}{The Breakthrough Propulsion Physics
website is http://www.lerc.nasa.gov/WWW/bpp/}

\bigskip
{\parindent 0.4truein \narrower \noindent
NASA is embarking on a new, small program called Breakthrough Propulsion
Physics to seek the ultimate breakthroughs in space transportation: (1)
Propelling a vehicle without propellant mass, (2) attaining the maximum
transit speeds physically possible, and (3) creating new energy production
methods to power such devices. Because such goals are beyond the
accumulated scientific knowledge to date, further advances in science are
sought,
specifically advances that focus on propulsion issues. Because such goals
are presumably far from fruition, a special emphasis of this program is to
demonstrate that near-term, credible, and measurable progress can be made.
This program, managed by Marc Millis of Lewis Research Center
(LeRC) represents the combined efforts of individuals from various NASA
centers, other government labs, universities and industry. This program
is supported by the Space Transportation Research Office of the Advanced Space
Transportation Program managed by Marshall Space Flight
Center (MSFC).

}

\bigskip\noindent
The first NASA BPP workshop was held in August 1997 to survey the territory
and assess
emerging physics concepts. Several invited presentations discussed the ZPF
vacuum
fluctuations, and this area of research was given a high priority in a
ranking process
carried out as part of the meeting (Millis 1998 and references therein). In
addition to
the proposed ZPF-inertia and ZPF-gravitation hypotheses, the possibility of
extracting
energy and of generating forces from the vacuum fluctuations were
discussed. It has been
shown that extracting energy from the vacuum does not violate the laws of
thermodynamics
(Cole and Puthoff 1993). As for ZPF-related forces, the recent measurements
of the Casimir
force by Lamoreaux (1997) are in agreement with theoretical predictions.
Real, macroscopic forces can be attributed to certain configurations of the
ZPF, such as
in a Casimir cavity. {\it We are proposing that inertia too is a Casimir-like
acceleration-dependent drag force.}

\bigskip\noindent
\centerline{\bf{NEWTON'S EQUATION OF MOTION: {\bf f}=m{\bf  a}}}
\bigskip\noindent
Physics recognizes the existence of four types of mass.
(1) {\it Inertial
mass:} the resistance to acceleration known as inertia, defined in Newton's
equation of motion,
${\bf f}=m{\bf a}$, and its relativistic generalization. (2) {\it Active
gravitational mass:} the ability of matter to attract other matter via
Newtonian gravitation, or, from the perspective of general relativity, the
ability to curve spacetime. (3) {\it Passive gravitational mass:} the
propensity of matter to respond to gravitational forces. (4) {\it Relativistic
rest mass:} the relationship of the mass of a body and the total energy available by
perfect annihilation of the mass in the body, that is expressed in the
$E=mc^2$ relation of special relativity. These are very different properties of matter,
yet for some reason they are quantitatively represented by the same parameter.
One can imagine a universe, for example, in which inertial mass,
$m_i$, and passive gravitational mass, $m_g$, were different\dots but then
objects would not all fall with the same acceleration in a gravitational field
and there would be no principle of equivalence to serve as the foundation of
general relativity. One can imagine a universe in which active and passive
gravitational mass were different\dots but then Newton's third law of equal and
opposite forces would be violated, and mechanics as we know it would be
impossible.

\bigskip\noindent
Consider inertial mass, $m_i$. Exert a certain force, {\bf f}, and measure
a resultant
acceleration, {\bf a}. Let this process take place under ideal conditions
of zero friction. A nearly perfect example --- excluding the very small residual atmospheric drag even at Shuttle altitudes --- would be the force exerted by
the Space Shuttle
engines and the acceleration of the Shuttle that results upon firing.
The inertial mass
is a scalar coefficient linking these two measureable processes {\bf f} and
{\bf a} (scalar
since the vectors {\bf f} and {\bf a} point in the same direction). However
since we
perceive a material object in the form of the Shuttle, we reify this $m_i$
coefficient and
attribute a property of mass to the object and then say that it is the mass
of the object
that causes the resistance to acceleration. That is to say, for a given
amount of
$m_i$ residing in the matter of an object it takes so much force to achieve
such a rate of
acceleration, which is embodied in
${\bf f}=m{\bf a}$. We thus attribute mass to all material objects.

\bigskip\noindent
It is important to keep in mind that the {\it actual direct measurement} of the
thing we call
inertial mass,
$m_i$, can only take place during acceleration\dots or deceleration which is simply
acceleration directed opposite to the existing velocity. We {\it assume} that an object
always possesses something called mass even when it is not accelerating, and proceed to
calculate the momentum,
$m_iv$, and the kinetic energy,
$m_iv^2/2$, of an object moving at constant velocity with respect to us. But
there can be no
direct evidence that an object possesses mass unless it is being accelerated.
The only way we can directly measure the momentum or the kinetic energy
that we calculate
is by bringing about a collision. But a collision necessarily involves
deceleration. It
makes for good bookkeeping to assume that an object always carries with it
a thing
called mass, yielding a certain momentum and kinetic energy, but this is
necessarily an
abstraction.

\bigskip\noindent
The momentum and kinetic energy depend upon relative motion, since no
velocity is
absolute. Move alongside an object and its momentum and kinetic energy
reduce to zero.
We argue that in a somewhat analogous fashion, $m_i$ is not something that
resides
innately in a material object, but rather that it is an electromagnetic
reaction force (per
unit acceleration) that springs into existence the instant an acceleration
occurs, and
disappears as soon as the acceleration stops. It is, precisely as defined
in Newton's {\bf
f}=m{\bf a}, a coefficient linking force and acceleration. It is a force
per unit
acceleration that arises electrodynamically.

\bigskip\noindent
This may be brought into sharper focus by considering Newton's third law.
Newton's third law states that
for every force there must be an equal and opposite reaction force, i.e. ${\bf
f}=-{\bf f}_r$. For stationary or static phenomena it is impossible to
even conceive of an alternative: If the right hand is pressing against the left
hand with force {\bf f}, then the left hand must press back against the right
hand with the equal and oppositely-directed reaction force, ${\bf f}_r$. How
could one hand press against the other without the other pressing back? It
would violate a fundamental symmetry, since who or what is to say which hand is
pressing and which is not. Thus for static or stationary situations the balance
of forces is the only imaginable circumstance.

\bigskip\noindent
If an agent exerts a force on a non-fixed object, experience tells us that a
reaction force also manifests against the agent. But why is this so? The
traditional explanation is that matter possesses inertial mass which by its
nature resists acceleration by pushing back upon the agent. The discovery
that we have made
is that, on the contrary, there is a very specific electromagnetic origin
for a reaction
force
${\bf f}_r$. Accelerated motion through the electromagnetic zero-point field
(ZPF) of the quantum vacuum results in a reaction force. If one analyses the
ZPF using Maxwell's equations of electrodynamics, one finds that ${\bf f}_r =
-m_{zp}{\bf a}$ where $m_{zp}$ is an electromagnetic parameter with units
of mass. An
electromagnetic reaction force (somewhat like a drag force) arises that
happens to be
proportional to acceleration. In other words, if one begins with
Maxwell's equations as applied to the ZPF, one finds from the laws of
electrodynamics that
${\bf f}_r = -m_{zp}{\bf a}$ and thus if one assumes that the
electrodynamic parameter
$m_{zp}$ really {\it is} the physical basis of mass,
Newton's third law of
equal and opposite forces, ${\bf f}=-{\bf f}_r$, results in a {\it
derivation} of
${\bf f}=m{\bf a}$ from the electrodynamics of the ZPF. That being the
case, one can, in
principle, dispense with the concept of inertial mass altogether. Matter,
consisting of
charged particles (quarks and electrons) interacts with the electromagnetic
ZPF and this
yields a reaction force whenever acceleration takes place and that is the
cause of inertia.

\bigskip\noindent
\centerline{\bf{THE ORIGIN OF THE ELECTROMAGNETIC ZERO-POINT FIELD}}
\bigskip\noindent
There are two views on the origin of the electromagnetic zero-point field
as embodied in
{\it Quantum Electrodynamics} (QED) and {\it Stochastic Electrodynamics} (SED)
respectively. The QED perspective is currently regarded as ``standard
physics'' and the
arguments go as follows. The Heisenberg uncertainty relation sets a
fundamental limit on
the precision with which conjugate quantities are allowed to be determined.
The two
principal conjugate pairs are position and momentum such that $\Delta x
\Delta p \ge \hbar/2$,
and energy and time such that $\Delta E \Delta t \ge \hbar/2$ where $\hbar$ is
Planck's constant, $h$, divided by $2\pi$. It
is a standard derivation in most textbooks on quantum mechanics to work out the
quantum version of a simple mechanical harmonic oscillator --- a mass on a
spring --- in this respect.

\bigskip\noindent
There are two non-classical results for a quantized harmonic oscillator.
First of all, the
energy levels are discrete and not continuous. By adding energy one can
increase the
amplitude of the oscillation, but only in units of $h\nu$, where $\nu$ is the
frequency in cycles per second. In other words, one can add or subtract
$E=nh\nu$ of
energy where $n\ge0$.
The second quantum effect stems from the fact that
if an oscillator were able to come completely to
rest,
$\Delta x$ would be zero and this would violate the $\Delta x \Delta p \ge \hbar/2$
limitation.
The result is that there is a minimum energy of $E=h\nu/2$, i.e. the
oscillator energy can
only take on the values $E=(n+1/2)h\nu$ which can never become zero since
$n$ cannot be
negative.

\bigskip\noindent
The argument is then made that the electromagnetic field is analogous to a
mechanical
harmonic oscillator since the electric and magnetic fields, {\bf E} and
{\bf B}, are modes of 
oscillating plane waves (see e.g. Loudon 1983). Each mode of oscillation of the
electromagnetic field has a minimum energy of $h\nu/2$. The volumetric density of modes
between frequencies $\nu$ and
$\nu+d\nu$ is given by the density of states function
$N_{\nu}d\nu=(8\pi\nu^2/c^3)d\nu$. Each state has a minimum $h\nu/2$
of energy,
and using this density of states function and this minimum energy that we call the zero-point
energy per state one can calculate the
ZPF spectral energy density:
$$\rho(\nu)d\nu={8\pi\nu^2 \over c^3} {h\nu
\over 2} d\nu .
\eqno(1)
$$

\bigskip\noindent
It is instructive to write the expression for zero-point spectral energy
density side by
side with blackbody radiation:

$$\rho(\nu,T)d\nu={8\pi\nu^2 \over c^3} \left( {h\nu \over e^{h\nu /kT} -1}
+{h\nu
\over 2}\right) d\nu .
\eqno(2)
$$
The first term (outside the parentheses) represents the mode density, and the
terms inside the parentheses are the average energy per mode of thermal
radiation at temperature $T$ plus the zero-point energy, $h\nu/2$, which has no
temperature dependence. Take away all thermal energy by formally letting $T$
go to zero, and one is still left with the zero-point term. The laws of quantum
mechanics as applied to electromagnetic radiation force the existence of a
background sea of zero-point-field (ZPF) radiation.

\bigskip\noindent
Zero-point radiation is a result of the application of quantum laws. It is
traditionally assumed in quantum theory, though, that the ZPF can for most practical
purposes be
ignored or subtracted away. The foundation of SED is the exact opposite. It
is assumed
that the ZPF is as real as any other electromagnetic field. As to its
origin, the
assumption is made that for some reason zero-point radiation just came with
the Universe.
The justification for this is that if one assumes that all of space is
filled with ZPF radiation, a
number of quantum phenomena may be explained purely on the basis of
classical physics
including the presence of background electromagnetic fluctuations provided
by the ZPF. The
Heisenberg uncertainty relation, in this view, becomes then not a result of
the existence
of quantum laws, but of the fact that there is a universal perturbing ZPF
acting on everything. The original motivation for developing SED was to see
whether the
need for quantum laws separate from classical physics could thus be
obviated entirely.

\bigskip\noindent
Philosophically, a universe filled --- for reasons unknown --- with a ZPF
but with only one
set of physical laws (classical physics consisting of mechanics and
electrodynamics), would
appear to be on an equal footing with a universe governed --- for reasons
unknown ---  by
two distinct physical laws (classical and quantum). In terms of physics,
though, SED
and QED are not on an equal footing, since SED has been successful in providing
a satisfactory alternative to only some quantum phenomena (although this
success does
include a classical ZPF-based derivation of the all-important blackbody
spectrum, cf.
Boyer 1984). Some of this is simply due to lack of effort: The ratio of
man-years devoted
to development of QED is several orders of magnitude greater than the expenditure
so far on SED.

\bigskip\noindent
\centerline{\bf{ACCELERATION AND THE DAVIES-UNRUH EFFECT}}
\bigskip\noindent
The ZPF spectral energy density of Eq. (1) would indeed be analogous to a
spatially
uniform constant offset that cancels out when considering energy fluxes.
However an
important discovery was made in the mid-1970's that showed that the ZPF
acquires special
characteristics when viewed from an accelerating frame.
In connection with radiation from evaporating black holes as proposed by
Hawking (1974), Davies (1975) and Unruh (1976) 
determined that a
Planck-like component of
the ZPF will arise in a uniformly-accelerated coordinate system having
constant proper
acceleration {\bf a} (where $|{\bf a}|=a$) with what amounts to an effective
``temperature''

$$T_a = {\hbar a \over 2 \pi c k} .  \eqno(3)$$

\smallskip\noindent
This ``temperature'' does not originate in emission from particles
undergoing thermal
motions.
\footnote{$^c$}{One suspects of course that there is a deep connection
between the fact
that the ZPF spectrum that arises in this fashion due to acceleration and
the ordinary
blackbody spectrum have identical form.}
As discussed by Davies, Dray and Manogue (1996):

\medskip
{\parindent 0.4truein \narrower \noindent
One of the most curious properties to be discussed in recent years is the
prediction that
an observer who accelerates in the conventional quantum vacuum of Minkowski
space will
perceive a bath of radiation, while an inertial observer of course
perceives nothing. In
the case of linear acceleration, for which there exists an extensive
literature, the
response of a model particle detector mimics the effect of its being
immersed in a bath of
thermal radiation (the so-called Unruh effect).

}
\medskip\noindent
This ``heat bath'' is a quantum phenomenon. The ``temperature'' is
negligible for most
accelerations. Only in the extremely large gravitational fields of black
holes or in
high-energy particle collisions  can this  become significant. This effect
has been
studied using both QED (Davies 1975, Unruh 1976) and in the SED formalism
(Boyer 1980). For the classical SED case it is found that the spectrum is
quasi-Planckian
in $T_a$. Thus for the case of no true external thermal radiation
$(T=0)$ but including this acceleration effect
$(T_a)$, equation (1) becomes

$$\rho(\nu,T_a)d\nu = {8\pi\nu^2 \over c^3}
\left[ 1 + \left( {a \over 2 \pi c \nu} \right) ^2 \right]
\left[ {h\nu \over 2} + {h\nu \over e^{h\nu/kT_a}-1} \right] d\nu , \eqno(4)$$

\smallskip\noindent where the acceleration-dependent pseudo-Planckian
component is placed
after the $h\nu/2$ term to indicate that except for extreme accelerations
(e.g. particle
collisions at high energies) this term is negligibly small.
While these additional acceleration-dependent terms do not show any spatial
asymmetry in
the expression for the ZPF spectral energy density, certain asymmetries do
appear when the
electromagnetic field interactions with charged particles are analyzed, or
when the
momentum flux of the ZPF is calculated. The ordinary plus $a^2$ radiation
reaction terms
in Eq.  (12) of HRP mirror the two leading terms in Eq. (4).

\bigskip\noindent
\centerline{\bf{THE ORIGIN OF INERTIA}}
\bigskip\noindent
Two independent approaches have demonstrated how a reaction force
proportional to
acceleration
$({\bf f}_r = -m_{zp} {\bf a})$ arises out of the properties of the ZPF.
The first
approach (HRP) was based upon a simplified model for how accelerated
idealized quarks and
electrons would interact with the ZPF. It identified the Lorentz force
arising from the
stochastically-averaged magnetic component of the ZPF, $<{\bf B}^{zp}>$, as
the basis of
${\bf f}_r$. The new approach (Rueda and Haisch 1998a, 1998b) 
considers
only the
relativistic transformations of the ZPF itself to an accelerated frame. We
find a non-zero
stochastically-averaged Poynting vector $(c/4 \pi)$ $<{\bf E}^{zp} \times {\bf
B}^{zp}>$
which leads immediately to a non-zero electromagnetic ZPF-momentum flux as
viewed by an
accelerating object. If the quarks and electrons in such an accelerating
object scatter
this asymmetric radiation, an acceleration-dependent reaction force ${\bf
f}_r$ arises. In
fact in this new analysis the ${\bf f}_r$ is the space-part of a
relativistic four-vector
so that the resulting equation of motion is not simply the classical ${\bf
f}=m{\bf a}$
expression, but rather the properly relativistic ${\cal F}=d{\cal P}/d\tau$
equation (that reduces exactly to ${\bf
f}=m{\bf a}$ for subrelativistic velocities).

\bigskip\noindent
In the first approach a specific ZPF-matter interaction is needed to carry
out the
analysis. We used a technique developed by Einstein and Hopf (1911) and
applied that to
idealized particles (partons, in the nomenclature of Feynman) treated as Planck
oscillators. In the second approach, no specific ZPF-matter interaction is
necessary for
the analysis. Any scattering or absorption process will yield a reaction
force on the
basis of a non-zero electromagnetic momentum flux. Presumably dipole
scattering of the ZPF
by fundamental charged particles is the appropriate representation, at
least to first
order, since that can be shown to be a detailed balance process in the
non-accelerated
case, i.e. dipole scattering by non-accelerated charged particles leaves
the ZPF spectrum
unchanged and isotropic (Puthoff 1989b). {\it In both approaches it is
assumed that the
level of interaction is that of quarks and electrons, which would account
for the inertial
mass of a composite neutral particle such as the neutron (udd).}

\bigskip\noindent
The expression for inertial mass in HRP {\it for an individual particle} is
$$m_{zp} ={\Gamma_z \hbar \omega_c^2 \over 2 \pi c^2},\eqno (5)$$
\medskip\noindent
where $\Gamma_z$ represents a damping constant for {\it zitterbewegung}
oscillations.
\footnote{$^d$}{In the Dirac theory of the electron, the velocity operator has eigenvalues
of $\pm c$.
The motion of an electron thus consists of two components: some average
motion specific to
a given physical circumstance plus an inherent highly oscillatory component
whose
instantaneous velocity is $\pm c$ which Schr\"odinger named {\it
zitterbewegung} (cf.
Huang 1952). The amplitude of this {\it zitterbewegung} oscillation is on
the order of the
Compton wavelength. From the perspective of the ZPF-inertia theory, the ZPF
can induce such
speed-of-light fluctuations since at this level the electron would be a
massless
point-charge. It is the Compton-wavelength size ``electron cloud'' that
acquires the
measured electron inertial mass of 512 keV in energy units via a
relationship like Eq. (5).
The
$\Gamma_e=6.25
\times 10^{-24}$ damping constant
governs the motion of the ``electron cloud'' whereas the $\Gamma_z$ applies
to the
internal {\it zitterbewegung}. This is an example of an SED interpretation
of an apparent
quantum phenomeon. The quantum size of the electron is its Compton
wavelength. The SED
interpretation would be one of a massless point charge driven by the ZPF to
oscillate at $\pm c$ within a Compton wavelength-size region of space. More on this is
extensively discussed in two articles by Rueda (1993).}
This is not to be confused with
$\Gamma_e=6.25
\times 10^{-24}$ s (Jackson 1975) which is used for macroscopic electron
oscillations in
ordinary radiation-matter interactions. $\Gamma_z$ is a free parameter and
$\Gamma_z \ne
\Gamma_e$. In Eq. (5)
$\omega_c$ represents an assumed cutoff frequency (in radians/s) for the
ZPF spectrum and
is also  a free parameter.

\bigskip\noindent
The expression for inertial mass in Rueda and Haisch (1998a, 
1998b) {\it
for an object with
volume $V_0$} is
$$m_{zp} = \left( {V_0 \over c^2} \int \eta(\omega) {\hbar \omega^3 \over 2
\pi^2
c^3} d\omega \right) = {V_0 \over c^2} \int \eta(\omega) \rho_{zp} \ d\omega.
\eqno (6)$$

\medskip\noindent
The interpretation of this is quite straightforward. The energy density of
the ZPF
(Eq. 1) written in terms of
$\omega(=2\pi\nu)$
is $\rho_{zp} d\omega =\hbar \omega^3 d\omega/2\pi^2 c^3$ which is the
second term in the integral. The dimensionless parameter $\eta(\omega)$
represents the
fraction of the ZPF flux scattered at each frequency. The total energy involved
``generating mass'' is determined by the volume of the object, $V_0$, and
the division by
$c^2$ converts the units to mass.

\bigskip\noindent
\centerline{\bf{CAN INERTIAL MASS BE ALTERED?}}
\bigskip\noindent
The mass of a proton in energy units is $\sim938$ MeV. A proton is composed
of two up (u)
quarks and one down (d) quark whose individual masses are $\sim5$ MeV for
the u, and
$\sim10$ Mev for the d. Thus the mass of the uud combination constituting
the proton is
about 50 times more massive than the sum of the parts. The same is true of
a neutron (udd)
whose mass is
$\sim940$ MeV. This is clearly a naive argument given the conceptual
uncertainty of what
``mass'' actually means for an individual quark which cannot exist in
isolation.
Nonetheless, taking this paradox at face value does offers a useful
perspective for
speculation.

\bigskip\noindent
The expression (Eq. 5) for $m_{zp}$ of an individual particle as derived by
HRP involves
two free parameters, $\Gamma_z$ and $\omega_c$. In HRP we assumed that
$\omega_c$ was some
cutoff frequency dictated either by an actual cutoff of the ZPF spectrum
(such as the
Planck frequency) or by a minimum size of a particle (such as the Planck
length). Let us
assume that in place of a cutoff frequency there is a resonance frequency
which is
specific to a given particle, call it $\omega_0$.

\bigskip\noindent
One can now imagine that a u-quark has a resonance $\omega_0$(u) yielding
$m_{zp}=5$ MeV
and that the d-quark has a different resonance $\omega_0$(d) yielding
$m_{zp}=10$ MeV
(assuming the same $\Gamma_z$). It would not be surprising that a bound
triad of quarks
such as the uud or the udd would have a radically different resonance as an
ensemble. The
resonance of a mechanical system bears no simple relationship to the
resonances of its
component parts. On this basis it would be easy to see how the same three
quarks could have
a totally different mass collectively than individually.

\bigskip\noindent
This same line of reasoning could be applied to the concept of mass defect.
The sum of the
masses of two protons plus two neutrons is greater than the mass of a He
nucleus. Again,
one can easily imagine the resonance of a group of 12 bound quarks in a He
nucleus being
different than the sum of the resonances of four groups of three bound quarks.

\bigskip\noindent
The advantage of this line of reasoning is that one does not have to
convert mass into
energy and vice versa. The quarks themselves can remain basically unchanged
entities,
whereas the resonances characterizing the interaction between the quark
ensemble and the
ZPF vacuum vary. This view would not be at odds with the conventional
interpretation that in going
from two free protons plus two free neutrons to one bound He nucleus there
is simply a
change in potential (binding) energy taking place. That interpretation
becomes one way to
``balance the books'' but the change in resonance would serve equally well,
yet without
the need to convert something material (mass) into something immaterial
(energy). One
would then interpret the energy released during fusion in terms of a
change in the
kinetic energy of the {\it zitterbewegung} motions of the quarks, which are
driven by the
underlying vacuum. In other words, change in mass becomes instead a change in the amount
of energy
involved in ZPF-quark interactions resulting from changes in resonance. The
energy
released in fusion would be coming from the ZPF.

\bigskip\noindent
We are suggesting that the mass of a particle is determined by a resonance
frequency,
$\omega_0$, and that the mass of a composite entity can be radically
different from the
sum of the individual masses because of changes in the resonances due to
binding forces. If
that proves to be the case, then one would also expect the mass of an
individual particle
to be variable if a change in resonance can be induced via external boundary
conditions. This would be somewhat analagous to the well-known ability to
change
spontaneous emission (by more than an order of magnitude) by effectively
placing an atom in
an appropriate electromagnetic cavity.

\bigskip\noindent
We view inertia as a property a particle obtains in relation with the vacuum medium in which
it is immersed. We suggest that if one could somehow modify that vacuum medium then the mass
of a particle or object in it would change. There is in nature an outstanding anticipatory
example of a very analogous feature that is well known. This is the so-called ``equivalent
mass'' or ``effective mass'' concept that conducting electrons and holes display when
immersed in the crystal lattice of a semiconductor. The effective mass parameter was
introduced long ago: see for example Smith (1961). If an external agent applies a force to an
electron in the conduction band or to a hole in the valence band, the inertia response
obtained is not at all the one we would expect for an ordinary electron in empty space, but
rather is quite different from it depending on the details of the particular crystal
structure of the semiconductor in which the electron (or hole) is immersed. This is why these
particles are called ``quasiparticles'' in this situation with the effective mass being the
parameter that characterizes their inertial properties inside the semiconductor medium.
{\it The inertial property of the quasiparticle is due to the complex detailed interaction
with the surrounding crystal lattice.} The effective mass is modified if the potentials in
the crystal structure change. Moreover if the crystal structure has some anisotropy, the
effective mass is no longer a scalar, but a tensor.

\bigskip\noindent
We can very reasonably expect that if the vacuum is modified, particularly at high energies,
then our proposed inertial mass will also be modified and in particular, if one can manage to
introduce an anisotropy in such vacuum by modifying the structure of the vacuum modes in an
anisotropic way, the inertial mass may display tensorial properties. Such an anisotropy is
not unthinkable: A Casimir cavity is precisely a structure that introduces an anistropy of
the ZPF mode structure. It, of course, primarily effects low energy modes. We speculate that
we can one day modify the vacuum modes distribution even at high energies (particularly at
some particle resonance or resonances if these exist) perhaps by means of strong fields.

\bigskip\noindent
Therefore in semiconductors, the response of an electron to a given force is quite
different from
one material to another. The ``effective mass'' of an electron in silicon
is larger than
in gallium arsenide, for example. Although not directly a ZPF-determined effect, it nonetheless provides a cogent example as to how particle masses can depend on environments to which they are strongly coupled. A similar effect has recently been
reported for particles
produced inside collisions between heavy nuclei. Experimental evidence was
reported by Wurm
for a change in the effective mass of the $\rho$-meson during a collision
as reported by Schewe and Stein (A.I.P. Bulletin No. 369). The bulletin
also states:
``According to Volker
Koch of Lawrence Berkeley Laboratory, this effect can take place for particles
inside any nuclear environment, from the most common atoms to superdense
neutron stars.''

\bigskip\noindent
\centerline{\bf{ZPF-INDUCED GRAVITATION}}
\bigskip\noindent
One of the first objections typically raised against the existence of a real
ZPF is that the mass equivalent of the energy embodied in Eq. (1) would
generate an enormous spacetime curvature that would shrink the universe to
microscopic size. The resolution of this dilemma lies in the principle of
equivalence. If inertia is an electromagnetic phenomenon involving interactions
between charge and the ZPF, then gravitation must be a similar phenomenon. The
mere existence of a ZPF would not necessarily generate gravitation or spacetime
curvature. Indeed, preliminary development of a conjecture of Sakharov 
(1968)
by Puthoff (1989a) indicates that the ZPF in and of itself {\it cannot} be a
source of gravitation (see also discussion in Haisch and Rueda [1997]).

\bigskip\noindent
Expressed in the simplest possible way, all matter at the level of quarks and
electrons is driven to oscillate ({\it zitterbewegung} in the terminology of
Schr\"odinger) by the ZPF. But every oscillating charge will generate its own
minute electromagnetic fields. Thus any particle will experience the
ZPF as modified ever so slightly by the fields of adjacent particles\dots and
that is gravitation! It is a kind of long-range van der Waals force.

\bigskip\noindent
Such a ZPF-based theory of gravitation is only in the exploratory stage at this
point. The Puthoff (1989a) analysis that resulted in the calculation of a
proper
Newtonian inverse-square law of attraction has since been shown to be
problematic, e.g. see Carlip (1993) and the reply by Puthoff (1993), also
Cole, Danley and
Rueda (1998). Moreover at this time
there is no accounting for the gravitational deflection of light other than
to invoke a
variable permittivity and permeability of the vacuum due to the presence of
charged
matter. However if it can be shown that the dielectric properties of the
vacuum can be
suitably modified by matter so as to bring about light deflection, this may
be a viable
alternative interpretation to spacetime curvature since light propagation
serves to define the metric.

\bigskip\noindent
\centerline{\bf{CONCLUSIONS}}
\bigskip\noindent
A concept has been proposed
that attempts to account for the inertia of matter as an electromagnetic
reaction force. A parallel gravitation concept along
lines conjectured by Sakharov (1968) also exists in preliminary form, and is
consistent with
the proposed origin of inertia as demanded by the principle of equivalence.
On the
basis of this ZPF-inertia concept, we can definitively rule out one
speculatively hypothesized propulsion mechanism: matter possessing negative
inertial mass, a concept originated by Bondi (1957) is
shown to be logically impossible. One cannot ``turn around'' the reaction
force an object
experiences upon accelerating into an oppositely directed ZPF momentum flux.
What you move into comes at you.

\bigskip\noindent
Is it proper to regard the ZPF as a real electromagnetic field?
The measurements by Lamoreaux (1997) of the Casimir force show excellent
agreement --- at the five percent level (much better than previous
experiments)
--- with theoretical predictions. One interpretation of the Casimir force is
that it represents the radiation pressure resulting from the exclusion of
certain ZPF modes in the cavity between the (uncharged) conducting plates
(Milonni, Cook
and Goggin 1988). There are alternate ways of looking at this (cf. Milonni
1994). We
suggest that it is fruitful at this stage to continue exploring the
ramifications of a
real-ZPF paradigm and that just as a real, measureable Casimir force results
upon construction of an uncharged parallel-plate condenser, so too does a real,
measureable reaction force result upon acceleration thereby creating the
inertial properties of matter.

\bigskip\noindent
\centerline{\bf{ACKNOWLEDGEMENTS}}
\bigskip\noindent
We acknowledge support of NASA contract NASW-5050 for this work. BH also
acknowledges the
hospitality of Prof. J. Tr\"umper and the Max-Planck-Institut where some of
these ideas
originated during several extended stays as a Visiting Fellow. AR
acknowledges many
stimulating discussions with Dr. D. C. Cole.

{

\bigskip
\parskip=0pt plus 2pt minus 1pt\leftskip=0.25in\parindent=-.25in

\centerline{\bf{REFERENCES}}

\medskip
Bondi, H. (1957), ``Negative mass within general relativity'' {\it Rev.
Modern Phys.},
Vol. 29, No. 3, 423.

\medskip
Boyer, T.H. (1980), ``Thermal effects of acceleration through random
classical radiation'',
{\it Phys. Rev. D}, Vol. 21. 2137.

\medskip
Boyer, T. H. (1984) ``Derivation of the blackbody radiation spectrum from
the equivalence
principle in classical physics with classical electromagnetic zero-point
radiation,'' {\it
Phys. Rev. D}, 29, 1096.

\medskip
Carlip, S. (1993), ``Comments on `Gravity as a zero-point fluctuation
force''', {\it Phys.
Rev. A}, Vol. 47, 3452.

\medskip
Cole, D.C. and Puthoff, H.E. (1993) ``Extracting Energy and Heat from the
Vacuum,'' {\it Phys. Rev. E}, 48, 1562.

\medskip
Cole, D.C., Danley, K. and Rueda, A. (1998), ``Further analysis on gravity
originating
from a zero-point force,'' in preparation.

\medskip
Davies, P.C.W. (1975), ``Scalar particle production in Schwarzschild and
Rindler
metrics,'' {\it J. Phys. A}, Vol. 8, 609.

\medskip
Davies, P.C.W., Dray, T. and Manogue, C. A. (1996), ``The Rotating Quantum
Vacuum,'' {\it
Phys. Rev. D}, 53, 4382.

\medskip
Einstein, A. and Hopf, L. (1910), ``\"Uber einen Satz der
Wahrscheinlichkeitsrechnung und seine Anwendung in der Strahlungstheorie'',
{\it Annalen
der Physik (Leipzig)}, Vol. 33, 1096; ``Statistische Untersuchung der
Bewegung eines
Resonators in einem Strahlungsfeld'', Vol. 33, 1105.

\medskip
Haisch, B. and Rueda, A. (1997), ``Reply to Michel's `Comment on Zero Point
Fluctuations
and the Cosmological Constant,'' {\it Astrophys. J.}, 488, 563.

\medskip
Haisch, B., Rueda, A. and Puthoff, H.E. (1994; HRP), ``Inertia as a zero-point
field Lorentz Force,'' {\it Phys. Rev. A}, Vol. 49, 678.

\medskip
Hawking, S. (1974), ``Black hole explosions?'' {\it Nature}, 248, 30.

\medskip
Huang, K. (1952), ``On the Zitterbewegung of the Dirac Electron,'' {\it Am.
J. Phys.}, 20,
479.

\medskip
Jackson, J.D. (1975), {\it Classical Electrodynamics}, (Wiley and Sons),
ch. 17.

\medskip
Lamoreaux, S.K. (1997) ``Demonstration of the Casimir Force in the 0.6 to 6
$\mu$m Range,'' {\it Phys. Rev. Letters}, 78, 5.

\medskip
Loudon, R. (1983), {\it The Quantum Theory of Light (2nd. ed).}, Oxford
Univ. Press,
chap. 4.

\medskip
Matthews, R. (1995), ``Nothing Like a Vacuum,'' {\it New Scientist}, Vol.
145, No. 1966,
p. 30.

\medskip
Millis, M. G. (1998), ``Breakthrough Propulsion Physics Workshop
Preliminary Results,''
in {\it Space Technology and Applications International Forum--1998},
CP420, (M. S.
El-Genk, ed.), DOE CONF-980103, p. 3.

\medskip
Milonni, P.W. (1994), {\it The Quantum Vacuum}, Academic Pres, chap. 1.

\medskip
Milonni, P.W., Cook, R.J. and Goggin, M.E. (1988), ``Radiation pressure
from the vacuum:
Physical interpretation of the Casimir force,'' {\it Phys. Rev. A}, 38, 1621.

\medskip
Puthoff, H.E. (1989a), ``Gravity as a zero-point fluctuation force,'' {\it
Phys. Rev. A},
Vol. 39, 2333.

\medskip
Puthoff, H.E. (1989b), ``Source of vacuum electromagnetic zero-point
energy,'' {\it Phys.
Rev. A}, Vol. 40, 4857.

\medskip
Puthoff, H.E. (1993), ``Reply to `Comment on Gravity as a zero-point
fluctuation force''',
{\it Phys. Rev. A}, Vol. 47, 3454.

\medskip
Rueda, A. (1993), ``Stochastic Electrodynamics with Particle Structure.'' Parts I and
II. {\it Found. Phys. Letters}, 6, 75; and 6, 193.

\medskip
Rueda, A. and Haisch, B. (1998a), ``Inertia as reaction of the vacuum to
accelerated
motion,'' {\it Physics Letters A}, Vol. 240, 115. (also
http://xxx.lanl.gov/abs/physics/9802031)

\medskip
Rueda, A. and Haisch, B. (1998b), ``Contribution to inertial mass by
reaction of the vacuum
to accelerated motion,'' {\it Foundations of Physics}, 28, 1057. (also
http://xxx.lanl.gov/abs/physics/9802030)

\medskip
Sakharov, A. (1968), ``Vacuum Quantum Fluctuations in Curved Space and the
Theory of Gravitation,'' {\it Soviet Physics - Doklady}, Vol. 12, No. 11, 1040.

\medskip
Smith, R.A. (1961), {\it Semiconductors}, Cambridge Univ. press, ch. 2.

\medskip
Unruh, W.G. (1976) ``Notes on black-hole evaporation,'' {\it Phys. Rev. D},
Vol. 14, 870.

\medskip
Vigier, J.-P. (1995), ``Derivation of Inertial Forces from the Einstein-de Broglie-Bohm
(E.d.B.B.) Causal Stochastic Interpretation of Quantum Mechanics,'' {\it Found. Phys.}, 25,
1461.

}

\bye